\documentclass[12pt]{article}
\usepackage{times}
\usepackage{geometry}
\usepackage{graphicx}
\usepackage[utf8]{inputenc}
\usepackage{enumitem,amssymb}
\usepackage{ragged2e}
\usepackage{multirow}

\newlist{thematic}{itemize}{8}
\setlist[thematic]{label=$\square$}
\usepackage{pifont}
\usepackage{natbib}

\newcommand{\lya}{Ly$\alpha$}%

\begin{document}

\def\aj{{AJ}}                   
\def\actaa{{Acta Astron.}}      
\def\araa{{ARA\&A}}             
\def\apj{{ApJ}}                 
\def\apjl{{ApJ}}                
\def\apjs{{ApJS}}               
\def\ao{{Appl.~Opt.}}           
\def\apss{{Ap\&SS}}             
\def\aap{{A\&A}}                
\def\aapr{{A\&A~Rev.}}          
\def\aaps{{A\&AS}}              
\def\azh{{AZh}}                 
\def\baas{{BAAS}}               
\def\bac{{Bull. astr. Inst. Czechosl.}} 
\def\caa{{Chinese Astron. Astrophys.}}   
\def\cjaa{{Chinese J. Astron. Astrophys.}}  
\def\icarus{{Icarus}}           
\def\jcap{{J. Cosmology Astropart. Phys.}} 
\def\jrasc{{JRASC}}             
\def\memras{{MmRAS}}            
\def\mnras{{MNRAS}}             
\def\na{{New A}}                
\def\nar{{New A Rev.}}          
\def\nat{{Nature}}      
\def\pra{{Phys.~Rev.~A}}        
\def\prb{{Phys.~Rev.~B}}        
\def\prc{{Phys.~Rev.~C}}        
\def\prd{{Phys.~Rev.~D}}        
\def\pre{{Phys.~Rev.~E}}        
\def\prl{{Phys.~Rev.~Lett.}}    
\def\pasa{{PASA}}               
\def\pasp{{PASP}}               
\def\pasj{{PASJ}}               

\raggedright
\huge
Astro2020 Science White Paper \linebreak

Tracing the formation history of galaxy clusters into the epoch of reionization \linebreak
\normalsize

\noindent \textbf{Thematic Areas:} \hspace*{60pt} 
$\boxtimes$ Galaxy Evolution  $\boxtimes$ Cosmology and Fundamental Physics \linebreak
  
\textbf{Principal Author:}

Name: Roderik Overzier$^{1,2}$ and Nobunari Kashikawa$^{3}$
 \linebreak						

Institution:\linebreak 
$^{1}$ Observat\'orio Nacional, Rua Jos\'e Cristino, 77. CEP 20921-400, S\~ao Crist\'ov\~ao, Rio de Janeiro-RJ, Brazil.\linebreak 
$^{2}$ Institute of Astronomy, Geophysics and Atmospheric Sciences, University of S\~ao Paulo, S\~ao Paulo-SP, Brazil. \linebreak
$^{3}$ Department of Astronomy, Graduate School of Science, The University of Tokyo, 7-3-1 Hongo, Bunkyo, Tokyo 113-0033, Japan
 \linebreak
Email: roderikoverzier@gmail.com and n.kashikawa@astron.s.u-tokyo.ac.jp 
 \linebreak

\justify

\textbf{Abstract: 
The large-scale overdensities of galaxies at $\mathbf{z\simeq2-7}$ known as ``protoclusters'' are believed to be the sites of cluster formation, and deep, wide survey projects such as the Large Synoptic Survey Telescope (LSST) and the Wide Field Infrared Survey Telescope (WFIRST) will deliver significant numbers of these interesting structures. Spectroscopic confirmation and interpretation of these targets, however, is still challenging, and will require wide-field multi-plexed spectroscopy on $>$20 m-class telescopes in the optical and near-infrared. In the coming decade, detailed studies of protoclusters will enable us, for the first time, to systematically connect these cluster progenitors in the early universe to their virialized counterparts at lower redshifts. This will allow us to address observationally the formation of brightest cluster galaxies and other cluster galaxy populations, the buildup of the intra-cluster light, the chemical enrichment history of the intra-cluster medium, and the formation and triggering of supermassive black holes in dense environments, all of which are currently almost exclusively approached either through the fossil record in clusters or through numerical simulations. Furthermore, at the highest redshifts ($\mathbf{z\simeq5-10}$), these large extended overdensities of star-forming galaxies are believed to have played an important role in the reionization of the universe, which needs to be tested by upcoming experiments. Theory and recent simulations also suggest important links between these overdensities and the formation of supermassive black holes, but observational evidence is still lacking. In this white paper we review our current understanding of this important phase of galaxy cluster history that will be explored by the next generation of large aperture groundbased telescopes GMT and TMT.}

\pagebreak
\section{Introduction}

The study of cluster formation, and the properties of galaxies as a function of environment in general, is a largely unexplored area of research at high redshifts and a key science driver for many new instruments and surveys. The period between cosmic dawn and cosmic noon, in particular, was fundamental for the formation of massive galaxy clusters, a process about which still very little is known from direct empirical evidence. Simulations show that cluster forming regions at $z\gtrsim2$ are characterized by large complexes of infalling dark matter halos with associated overdensities of gas and galaxies that can offer important constraints on the various processes relevant to galaxy clusters, including the infall of matter and galaxies along cosmic filaments, galaxy interactions and mergers in dense regions, co-evolution of supermassive black holes and massive galaxies, the build-up of the brightest cluster galaxy (BCG) and its intra-cluster light envelope, the roles of star formation, outflows and quenching, and the origin and its metal enrichment of the intra-cluster medium (ICM). Despite the enormous parameter space left to be explored, these dense regions in the early universe are rare and still a challenge to study observationally \citep{overzier16}.

Our current knowledge on the formation history of galaxy clusters mainly comes from two areas of research. In principle, galaxy clusters in the local universe up to about cosmic noon contain a detailed fossil record of their entire formation history. This information mainly comes in the form of the properties of their galaxy populations (e.g., morphologies and stellar populations), properties of the intra-cluster gas (e.g., temperature, luminosity and metal abundances) and the properties of their dark matter (sub)structure (e.g., mass and density profile). However, interpretation of this fossil record is not straightforward, as the massive galaxy population that dominates galaxy clusters formed long before they became part of the cluster, and because the cluster virialization process largely removes information about the initial conditions of structure before and during the collapse. Therefore, we have mainly relied on cosmological numerical simulations that reproduce the properties of clusters in observations as a way to understand and test our models of cluster formation and evolution. While examples of massive, virialized or nearly virialized clusters have been found up to $z\approx2$, these systems often show properties that are remarkably similar to those of massive clusters at lower redshifts in a way that obfuscates their origin \citep[e.g.][]{newman14,brodwin16}. A complementary approach is to search for large-scale, pre-virialized overdense regions in the distribution of galaxies or gas at high redshifts (often at $z\gtrsim2$) that are good candidates of clusters-in-formation. These regions should be both sufficiently large and overdense in order to make sure that they will collapse into a cluster-sized, virialized object by the present-day. Such objects are now being found in relatively large numbers, which is providing direct empirical evidence on the earlier phase during which the clusters and their galaxies were actually forming. 

\begin{figure}[t]
\begin{center}
\includegraphics[width=0.8\textwidth]{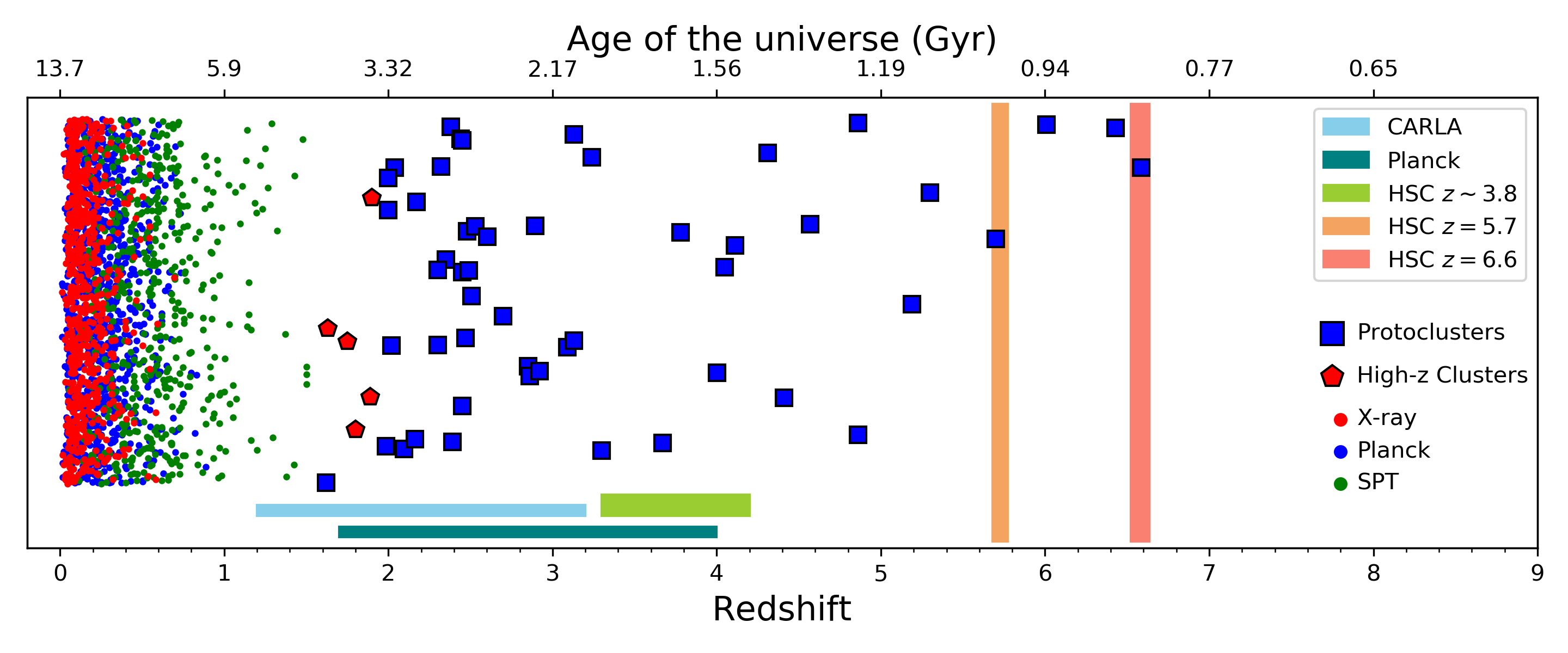}
\end{center}
\caption{\label{fig:cone}{\bf Overview of protoclusters found to date. Horizontal shaded regions show the approximate redshift ranges probed by the large numbers of (candidate) clusters and protoclusters from the Clusters Around Radio Loud AGN (CARLA) project \citep{wylezalek13}, the Planck-detected protocluster candidates \citep{planck15} and the HSC selected protoclusters at $\mathbf{z\sim4}$ \citep{toshikawa18}. Vertical shaded regions indicate the approximate redshift ranges probed by the \lya\ narrowbands of the HSC SSP at $\mathbf{z\approx5.7}$ and $\mathbf{z\approx6.6}$ \citep[e.g.][]{higuchi18,harikane19}. Data on low redshift clusters selected in X-ray, Planck and SPT was taken from \citet{bleem15}. The vertical axis does not contain any information.}}
\end{figure}

\section{A census of cluster formation: from butterfly collecting to evolutionary biology}

After two decades of finding unique examples of cluster progenitors (``protoclusters") rather sporadically through a variety of methods and at a wide range of redshifts (about $\sim$40 were known at $z\simeq2-6$ until recently), this area of research is reaching maturity. Deep, wide multi-wavelength sky surveys are starting to reach both the necessary depth and areas large enough to detect statistical numbers of the rare protoclusters in a more uniform and systematic manner (see Figure 1). Data from the HyperSuprimeCam (HSC) Subaru Systematic Program (SSP) was recently used to select $\sim200$ protoclusters at $z\sim3.8$ based on high density peaks in the sky distribution of $g$-band dropout galaxies \citep[Figure 2;][]{toshikawa18}, exceeding the total number of protoclusters known in this redshift range prior to this study by more than a factor of 10. HSC SSP narrow-band filters sensitive to redshifted \lya\ furthermore resulted in a large number of new candidate systems at $z\sim6$, several of which were confirmed spectroscopically \citep{higuchi18,harikane19}. Other sources of statistical samples of protoclusters are the Clusters Around Radio-Loud AGN (CARLA) survey, which has shown that many powerful radio-loud AGN are surrounded by galaxy excesses \citep{wylezalek13}, the Planck survey which has revealed a large number of unresolved ``cold'' sources associated with large overdensities of dust-obscured star-forming galaxies at high redshift \citep{planck15}, the VIMOS Ultra Deep Survey (VUDS) which has found several bonafide protoclusters owing to its dense spectroscopic sampling of the cosmic web \citep{cucciati14,cucciati18,lemaux14,lemaux18}. Other recent examples of protoclusters include several systems identified as highly clustered regions of dusty star-forming galaxies in data from the Herschel Space Observatory and the South Pole Telescope \citep[e.g.][]{miller18,gomez19}, and large protocluster systems detected in hydrogen gas absorption studies \citep[e.g.][]{lee16,cai17,hayashino19}. The data on all these targets span a wide range in redshift and likely include progenitors of present-day clusters with a range of cluster mass and evolutionary states. During the next decade, systematic multiwavelength followup of  these systems and comparison with simulations will provide a detailed picture of cluster formation from $z=0$ to the epoch of reioniation (EoR).

\begin{figure}[t]
\begin{center}
\includegraphics[width=0.9\textwidth]{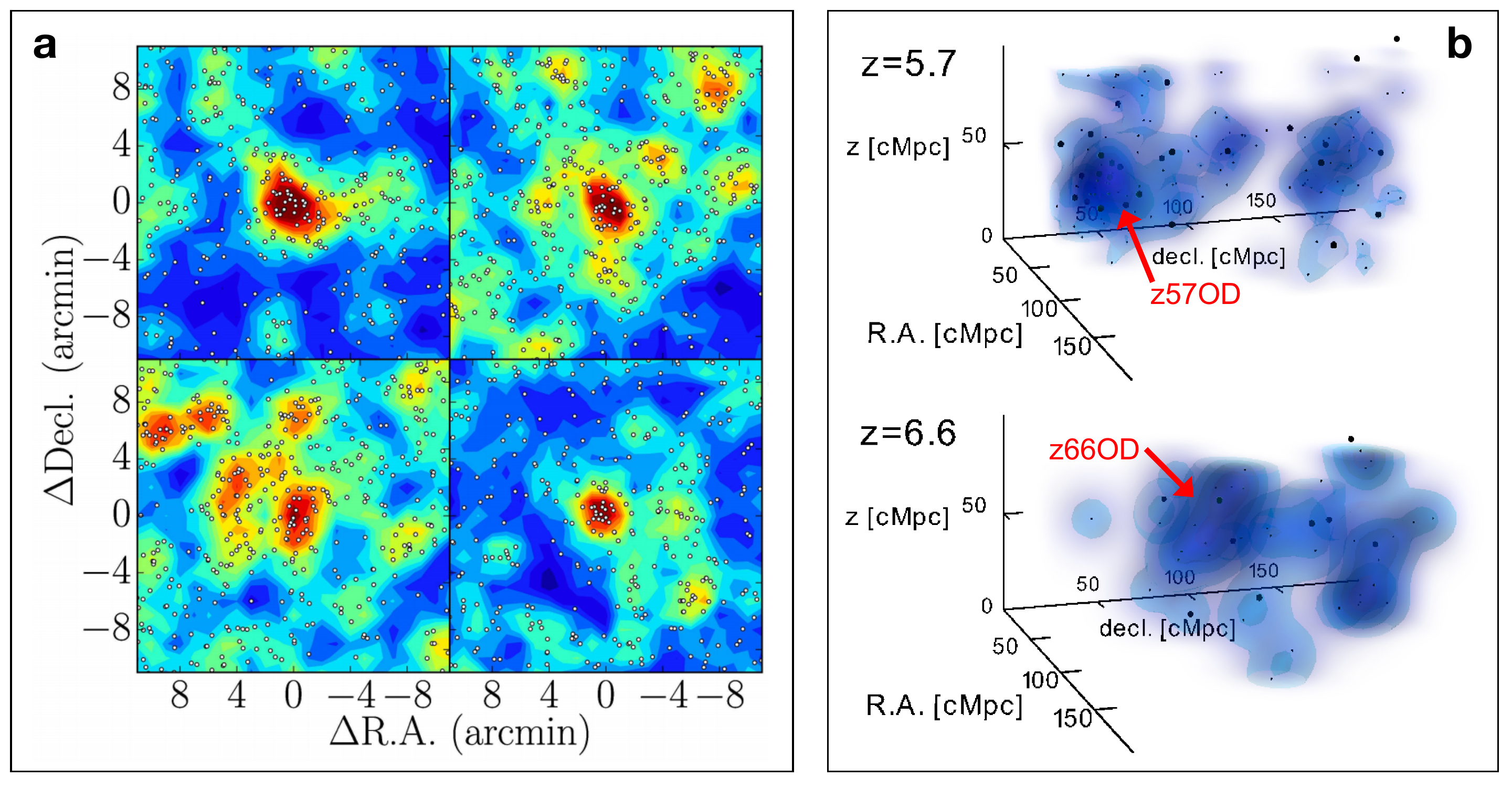}
\end{center}
\caption{\label{fig:examples}{\bf Examples of recently discovered protoclusters at $\mathbf{z\sim4}$ \citep[panel (a); from][]{toshikawa18} and at $\mathbf{z\sim6-7}$ \citep[panel (b), from][]{harikane19}.}}
\end{figure}

\section{Protoclusters and the epoch of reionization}

The epoch of reionization is a focus area of cosmology and galaxy evolution. Because the values of the cosmological parameters together with structure formation set the growth rate of the first stars and galaxies that ionized the intergalactic medium, studying the process of reionization is analogous to studying the formation of the first structures. There is mounting evidence that protoclusters have played an important role in this process. Protoclusters represent significant density fluctuations of dark matter, gas and galaxies up to very large scales \citep[tens of Mpc;][]{chiang13}. Because galaxy formation in these dense regions preceeded that in lower density regions and because they contained higher densities of ionizing-photon producing objects per unit volume, the contribution from protoclusters to the total ionizing-photon rate during the EoR may have been substantial. \citet{chiang17} showed using simulations that at $z\sim6-10$, protocluster galaxies are responsible for around 30--50 \% of the cosmic star formation rate density even though they occupy only $\sim$6 \% of the cosmic volume. Although the exact contribution from these objects to the cosmic ionized fraction and the ionized bubble sizes around protoclusters as a function of redshift remain to be determined, these numbers suggest that their contribution could be substantial. 

Several examples of such objects have already been discovered at $z\simeq6-8$ \citep[e.g., see Figure 2;][]{ouchi05,trenti12,toshikawa14,ishigaki16,higuchi18,jiang18,harikane19}. Recently, a large systematic survey of protoclusters in the EoR was performed by \citet{higuchi18}, who selected several tens of candidate structures at $z=5.7$ and $z=6.6$ based on overdensities of LAEs in data from the HSC SSP.  
These structures could be used to study their impact on reionization in various ways. It is expected that high density regions contain star-forming galaxies with larger-than-average equivalent widths of \lya\ because the escape of \lya\ photons is facilitated by the large ionized bubbles around the protoclusters \citep{inoue18,higuchi18}. Likewise, the shape of the typical \lya\ profiles as a function of redshift and/or overdensity can be used to probe any changes in the IGM opacity \citep[e.g.][]{kashikawa11}. Because the \lya\ emission is often offset with respect to the systemic velocity of the galaxies, it is important to determine accurate redshifts based on the fainter interstellar absorption lines in the rest-frame UV or strong nebular emission lines in the rest-frame optical. Protoclusters can furthermore be used as highly efficient probes of the process of Lyman continuum (LyC) leakage that is crucial to the EoR. Because protoclusters have high surface densities of galaxies at a known redshift, a single pointing with a multi-plexed spectrograph will result in multiple high signal-to-noise spectra that can be searched for LyC emission \citep[e.g.][]{mostardi13}. By studying the measured escape fractions as a function of the physical properties of the galaxies, the physics that allowed LyC escape to occur can be probed \citep[e.g.][]{borthakur14}. By determining the star formation histories of galaxies in protoclusters directly after or during the end of the EoR, it may even be possible to infer their contribution to the ionizing photon budget at earlier times. The spatial distribution of galaxies in protoclusters combined with maps of the neutral hydrogen distribution from the Square Kilometer Array (SKA) will determine the topology of these structures during the EoR.  

Finally, there may exist important links between the first overdense regions and the formation of supermassive black holes (SMBHs). During the past decade, many authors have searched for structures of galaxies near luminous quasars at $z\gtrsim5$ with a wide range in outcomes \citep[e.g.][]{kashikawa07,overzier09,mazzucchelli17,balmaverde17,champagne18,ota18}. The strongest evidence to date that suggests that quasars at $z\gtrsim6$ are located in the densest regions of the cosmic web comes from models and simulations, but the empirical evidence is still lacking. Recent simulations have suggested that the presence of dense star-forming regions, such as protoclusters, may even be required in order to facilitate the direct collapse of massive SMBH seeds during the EoR, in part through the enhanced Lyman-Werner background generated by these dense regions \citep{wise19}. On the other hand, the strong photoionization effects from luminous quasars may suppress gas cooling in surrounding halos, which can be tested by comparing the luminosity functions of young star-forming objects in their vicinity with those in the field \citep[e.g.][]{uchiyama19}. There are thus strong additional motivations to search in and around these dense regions for early SMBH activity, and vice versa. The ongoing discovery and detailed study of dense protocluster regions on one hand and  luminous or massive galaxies and black holes in the EoR should be able to shed light on the relation between SMBH formation and the densest regions in the early universe.

\section{Open questions and recommendations}

How the aforementioned protoclusters exactly relate to the galaxy clusters at lower redshifts still remains an open question in most cases. Likewise, their contribution to the cosmic star formation rate and reionization needs to be determined through observations. Observations of galaxies in galaxy clusters and theoretical models have demonstrated that galaxy evolution depends on, or correlates with, the environment \citep[e.g.,][]{delucia07,thomas10,behroozi13,contini16,laigle18}. Most of the stars in massive cluster galaxies formed long before cluster virialization, and especially BCGs may have formed nearly half of their stars by $z\sim5$. Some authors have found differences in the stellar mass distributions, ages of the stellar populations and/or AGN fractions in protoclusters compared to the field \citep[e.g.,][]{steidel05,cooke14,macuga18}. The overdense environments could also lead to differences in the gas-phase metal abundances of the galaxies, for example due to the accretion of gas that is metal-enhanced compared to lower density regions, more efficient stripping of gas in dense environments, or dilution of metallicities due to efficient inflows of gas \citep[e.g.][]{shimakawa15,valentino15,overzier16}. Outflows from starburst galaxies during the protocluster phase were important for the enrichment of the present-day ICM \citep{biffi18}. Some protoclusters contain particularly massive central galaxies or dense concentrations of galaxies that are good candidates of proto-BCGs \citep[e.g.][]{pentericci97,kubo16,miller18}, but a clear picture of BCG assembly has yet to emerge. For massive clusters, many of these important events took place at $z\gtrsim2$, making the protocluster phase in principle one of most informative phases about cluster formation. 

Current studies of protoclusters are hampered by at least three important factors: limited spectroscopic determinations of membership, limited multi-wavelength coverage, and small number statistics. The combination of optical/near-infrared observations with the Giant Magellan Telescope (GMT) and the Thirty Meter Telescope (GMT), mid-infrared observations with the James Webb Space Telescopes (JWST), and the Atacama Large Millimeter Array (ALMA) will be particularly powerful to constrain the physical properties of protocluster galaxies and the mass and dynamical states of selected systems. Protocluster galaxies, which are more strongly clustered than field galaxies, are good targets for planned multi-object spectrographs on $>$20 meter class telescopes as they can take full advantage of the multiplexity. The large collecting area of these telescopes will furthermore increase the speed at which protocluster candidates can be confirmed, increasing the sample size and allowing basic structural properties to be determined (e.g., redshift, overdensity, velocity dispersion and mass). The main diagnostic features in the rest-frame UV from the Lyman limit to [MgII]$\lambda\lambda$2796,2803 fall in the observed optical/near-infrared for targets between $z\sim3$ and $z\sim7$. The main optical emission lines up to the [SII]$\lambda\lambda$6717,6731 doublet will be accessible for protoclusters at $z\lesssim2.6$, and the spectral range around the Balmer break at $z\lesssim5$. Combined with other multi-wavelength data, this will allow the measurement of stellar populations, stellar mass and age, LyC and \lya\ escape fractions, extinction, outflows, metallicity and interstellar medium properties. Estimates for the velocity dispersions and current and final masses of the protoclusters can then be used to place them on different evolutionary tracks appropriate for their mass scale \citep[e.g.,][]{chiang13,miller18,toshikawa18}. This will allow us to compare the properties of their galaxies with those of their field counterparts at the same redshift in order to study environmental trends or look for evidence of assembly bias. 

Typical protoclusters measure several to about ten arcminutes in diameter (see Figure 2), making them particularly convenient targets for medium-resolution optical multi-object spectrographs such as GMACS (possibly extended to larger areas with MANIFEST) on the GMT and WFOS on the TMT. These telescopes will also allow a dense sampling of the gas distribution associated with protoclusters through cross-correlation with \lya\ absorption features detected in the continuum spectra of faint background galaxies. Diffraction-limited images and multi-object or integral field spectroscopy of selected high density regions or particular individual objects in the near-infrared can be studied with GMT (GMTIFS) and TMT (IRIS, IRMS and IRMOS). If any quasar activity is present in or near protoclusters, virial masses of their SMBHs can be estimated from the MgII line and the continuum flux at 3000 \AA. The properties of \lya\ will inform us about the ionization state of the densest environments, which are believed to be major ionizing sources during the EoR.


\clearpage

\begin{thebibliography}{}
\bibitem[Balmaverde et al.(2017)]{balmaverde17} Balmaverde, B., Gilli, R., Mignoli, M., et al.\ 2017, \aap, 606, A23 
\bibitem[Behroozi et al.(2013)]{behroozi13} Behroozi, P.~S., Wechsler, R.~H., \& Conroy, C.\ 2013, \apj, 770, 57 
\bibitem[Biffi et al.(2018)]{biffi18} Biffi, V., Planelles, S., Borgani, S., et al.\ 2018, \mnras, 476, 2689 
\bibitem[Bleem et al.(2015)]{bleem15} Bleem, L.~E., Stalder, B., de Haan, T., et al.\ 2015, \apjs, 216, 27 
\bibitem[Borthakur et al.(2014)]{borthakur14} Borthakur, S., Heckman, T.~M., Leitherer, C., \& Overzier, R.~A.\ 2014, Science, 346, 216 
\bibitem[Brodwin et al.(2016)]{brodwin16} Brodwin, M., McDonald, M., Gonzalez, A.~H., et al.\ 2016, \apj, 817, 122 
\bibitem[Cai et al.(2017)]{cai17} Cai, Z., Fan, X., Bian, F., et al.\ 2017, \apj, 839, 131 
\bibitem[Champagne et al.(2018)]{champagne18} Champagne, J.~B., Decarli, R., Casey, C.~M., et al.\ 2018, \apj, 867, 153 
\bibitem[Chiang et al.(2013)]{chiang13} Chiang, Y.-K., Overzier, R., \& Gebhardt, K.\ 2013, \apj, 779, 127 
\bibitem[Chiang et al.(2017)]{chiang17} Chiang, Y.-K., Overzier, R.~A., Gebhardt, K., \& Henriques, B.\ 2017, \apjl, 844, L23 
\bibitem[Contini et al.(2016)]{contini16} Contini, E., De Lucia, G., Hatch, N., Borgani, S., \& Kang, X.\ 2016, \mnras, 456, 1924 
\bibitem[Cooke et al.(2014)]{cooke14} Cooke, E.~A., Hatch, N.~A., Muldrew, S.~I., Rigby, E.~E., Kurk, J.~D.\ 2014, \aap, 570, A16 
\bibitem[Cucciati et al.(2014)]{cucciati14} Cucciati, O., Zamorani, G., Lemaux, B.~C., et al.\ 2014, \aap, 570, A16 
\bibitem[Cucciati et al.(2018)]{cucciati18} Cucciati, O., Lemaux, B.~C., Zamorani, G., et al.\ 2018, \aap, 619, A49 
\bibitem[De Lucia \& Blaizot(2007)]{delucia07} De Lucia, G., \& Blaizot, J.\ 2007, \mnras, 375, 2 
\bibitem[G{\'o}mez-Guijarro et al.(2019)]{gomez19} G{\'o}mez-Guijarro, C., Riechers, D.~A., Pavesi, R., et al.\ 2019, \apj, 872, 117 
\bibitem[Harikane et al.(2019)]{harikane19} Harikane, Y., Ouchi, M., Ono, Y., et al.\ 2019, arXiv:1902.09555 
\bibitem[Hayashino et al.(2019)]{hayashino19} Hayashino, T., Inoue, A.~K., Kousai, K., et al.\ 2019, \mnras, 484, 5868 
\bibitem[Higuchi et al.(2018)]{higuchi18} Higuchi, R., Ouchi, M., Ono, Y., et al.\ 2018, arXiv:1801.00531 
\bibitem[Inoue et al.(2018)]{inoue18} Inoue, A.~K., Hasegawa, K., Ishiyama, T., et al.\ 2018, \pasj, 70, 55 
\bibitem[Ishigaki et al.(2016)]{ishigaki16} Ishigaki, M., Ouchi, M., \& Harikane, Y.\ 2016, \apj, 822, 5 
\bibitem[Jiang et al.(2018)]{jiang18} Jiang, L., Wu, J., Bian, F., et al.\ 2018, Nature Astronomy, 2, 962 
\bibitem[Kashikawa et al.(2007)]{kashikawa07} Kashikawa, N., Kitayama, T., Doi, M., et al.\ 2007, \apj, 663, 765 
\bibitem[Kashikawa et al.(2011)]{kashikawa11} Kashikawa, N., Shimasaku, K., Matsuda, Y., et al.\ 2011, \apj, 734, 119 
\bibitem[Kubo et al.(2016)]{kubo16} Kubo, M., Yamada, T., Ichikawa, T., et al.\ 2016, \mnras, 455, 3333 
\bibitem[Lee et al.(2016)]{lee16} Lee, K.-G., Hennawi, J.~F., White, M., et al.\ 2016, \apj, 817, 160 
\bibitem[Laigle et al.(2018)]{laigle18} Laigle, C., Pichon, C., Arnouts, S., et al.\ 2018, \mnras, 474, 5437 
\bibitem[Lemaux et al.(2014)]{lemaux14} Lemaux, B.~C., Cucciati, O., Tasca, L.~A.~M., et al.\ 2014, \aap, 572, A41 
\bibitem[Lemaux et al.(2018)]{lemaux18} Lemaux, B.~C., Le F{\`e}vre, O., Cucciati, O., et al.\ 2018, \aap, 615, A77 
\bibitem[Macuga et al.(2018)]{macuga18} Macuga, M., Martini, P., Miller, E.~D., et al.\ 2018, arXiv:1805.06569 
\bibitem[Madau \& Dickinson(2014)]{madau14} Madau, P., \& Dickinson, M.\ 2014, \araa, 52, 415 
\bibitem[Mazzucchelli et al.(2017)]{mazzucchelli17} Mazzucchelli, C., Ba{\~n}ados, E., Decarli, R., et al.\ 2017, \apj, 834, 83 
\bibitem[Miller et al.(2018)]{miller18} Miller, T.~B., Chapman, S.~C., Aravena, M., et al.\ 2018, \nat, 556, 469 
\bibitem[Mostardi et al.(2013)]{mostardi13} Mostardi, R.~E., Shapley, A.~E., Nestor, D.~B., et al.\ 2013, \apj, 779, 65 
\bibitem[Newman et al.(2014)]{newman14} Newman, A.~B., Ellis, R.~S., Andreon, S., et al.\ 2014, \apj, 788, 51 
\bibitem[Ota et al.(2018)]{ota18} Ota, K., Venemans, B.~P., Taniguchi, Y., et al.\ 2018, \apj, 856, 109 
\bibitem[Ouchi et al.(2005)]{ouchi05} Ouchi, M., Shimasaku, K., Akiyama, M., et al.\ 2005, \apjl, 620, L1 
\bibitem[Overzier et al.(2009)]{overzier09} Overzier, R.~A., Guo, Q., Kauffmann, G., et al.\ 2009, \mnras, 394, 577 
\bibitem[Overzier(2016)]{overzier16} Overzier, R.~A.\ 2016, \aapr, 24, 14 
\bibitem[Pentericci et al.(1997)]{pentericci97} Pentericci, L., Roettgering, H.~J.~A., Miley, G.~K., Carilli, C.~L., \& McCarthy, P.\ 1997, \aap, 326, 580 
\bibitem[Planck Collaboration et al.(2015)]{planck15} Planck Collaboration, Aghanim, N., Altieri, B., et al.\ 2015, \aap, 582, A30 
\bibitem[Shimakawa et al.(2015)]{shimakawa15} Shimakawa, R., Kodama, T., Tadaki, K.-i., et al.\ 2015, \mnras, 448, 666 
\bibitem[Steidel et al.(2005)]{steidel05} Steidel, C.~C., Adelberger, K.~L., Shapley, A.~E., Erb, D.~K., Reddy, N.~A., Pettini, M.\ 2005, \apj, 626, 44 
\bibitem[Thomas et al.(2010)]{thomas10} Thomas, D., Maraston, C., Schawinski, K., Sarzi, M., \& Silk, J.\ 2010, \mnras, 404, 1775 
\bibitem[Toshikawa et al.(2014)]{toshikawa14} Toshikawa, J., Kashikawa, N., Overzier, R., et al.\ 2014, \apj, 792, 15 
\bibitem[Toshikawa et al.(2018)]{toshikawa18} Toshikawa, J., Uchiyama, H., Kashikawa, N., et al.\ 2018, \pasj, 70, S12 
\bibitem[Trenti et al.(2012)]{trenti12} Trenti, M., Bradley, L.~D., Stiavelli, M., et al.\ 2012, \apj, 746, 55 
\bibitem[Uchiyama et al.(2019)]{uchiyama19} Uchiyama, H., Kashikawa, N., Overzier, R., et al.\ 2019, \apj, 870, 45 
\bibitem[Valentino et al.(2015)]{valentino15} Valentino, F., Daddi, E., Strazzullo, V., et al.\ 2015, \apj, 801, 132 
\bibitem[Wise et al.(2019)]{wise19} Wise, J.~H., Regan, J.~A., O'Shea, B.~W., et al.\ 2019, \nat, 566, 85 
\bibitem[Wylezalek et al.(2013)]{wylezalek13} Wylezalek, D., Galametz, A., Stern, D., et al.\ 2013, \apj, 769, 79 
\end{thebibliography}
\end{document}